\documentclass{emulateapj}

\newcommand{\kms}{\mbox{$\>{\rm km\, s^{-1}}$}}
\newcommand{\pc}{\>{\rm pc}}
\newcommand{\kpc}{\mbox{$\>{\rm kpc}$}} 
\newcommand{\Gyr}{\mbox{$\>{\rm Gyr}$}}

\newcommand{\Msun}{\>{\rm M_{\odot}}}
 
\newcommand{\vgsr}{\mbox{$V_{\rm GSR}$}} 
\newcommand{\sgsr}{\mbox{$\sigma_{\rm GSR}$}} 
\newcommand\degrees{^\circ}
\newcommand{\avg}[1]{\mbox{$\left<{#1}\right>$}}

\newcommand{\tm}[1]{\mbox{$t_{#1}$}}

\def\araa{ARAA}
\def\aj{AJ}
\def\apj{ApJ}
\def\apjl{ApJ}
\def\apjs{ApJS}
\def\aap{A\&A}
\def\mnras{MNRAS}
\def\pasj{PASJ} 
\def\nat{Nature}

\def\aaps{A\&AS}

\def\etal{{et al.}}

\def\ie{{\it i.e.}}

\slugcomment{Draft to ApJ} 
\lefthead{Debattista \etal} 

\righthead{A Kiloparsec-Scale Nuclear Stellar Disk in the Milky Way}

\begin{document}   

\title{A Kiloparsec-Scale Nuclear Stellar Disk in the Milky Way as a
  Possible Explanation of the High Velocity Peaks in the Galactic
  Bulge}

\author{Victor P. Debattista\altaffilmark{1}, Melissa
  Ness\altaffilmark{2}, Samuel W.F. Earp\altaffilmark{1}, \& David
  R.Cole\altaffilmark{1}}

\altaffiltext{1}{Jeremiah Horrocks Institute, University of Central
  Lancashire, Preston PR1 2HE, UK}
\altaffiltext{2}{Max-Planck-Institut f\"ur Astronomie, K\"onigstuhl
  17, D-69117 Heidelberg, Germany}

\begin{abstract}

The Apache Point Observatory Galactic Evolution Experiment has
measured the stellar velocities of red giant stars in the inner Milky
Way.  We confirm that the line of sight velocity distributions
(LOSVDs) in the mid-plane exhibit a second peak at high velocities,
whereas those at $|b| = 2\degrees$ do not.  We use a high resolution
simulation of a barred galaxy, which crucially includes gas and star
formation, to guide our interpretation of the LOSVDs.  We show that
the data are fully consistent with the presence of a thin, rapidly
rotating, nuclear disk extending to $\sim 1 \kpc$.  This nuclear disk
is orientated perpendicular to the bar and is likely to be composed of
stars on x2 orbits.  The gas in the simulation is able to fall onto
such orbits, leading to stars populating an orthogonal disk.

\end{abstract}

\keywords{Galaxy: bulge --- Galaxy: evolution --- Galaxy: formation
  --- Galaxy: kinematics and dynamics --- Galaxy: stellar content}

%
%

\section{Introduction}
\label{sec:intro}

Detections of high Galactic standard-of-rest velocity (\vgsr) peaks in
the Apache Point Observatory Galactic Evolution Experiment (APOGEE)
commissioning data were reported by \cite{nidever+12} across all
fields at $4\degrees \leq l \leq 14\degrees$ and $-2\degrees \leq b
\leq 2\degrees$.  \citet{nidever+12} interpreted the high-\vgsr\ peaks
as being due to stars in the Galactic bar. However, the peaks are not
statistically significant in a number of fields \citep{li+14} and no
high-\vgsr\ peaks were found at negative longitudes in the Bulge
Radial Velocity Assay (BRAVA), at $b$ $\sim = -4 \degrees$
\citep{kunder+12}.  Additionally, no high-\vgsr\ peaks can be found in
pure $N$-body models \citep{li+14}.  \citet{molloy+15} demonstrated
that resonant (2:1 and higher order) orbits, viewed on their own, were
able to generate high-\vgsr\ peaks.  \citet{aumerschoenrich15}
proposed that such resonant orbits are populated by young stars
recently trapped by the bar; they argued that the APOGEE selection
function is biased toward such young stars.

Bars have been implicated in building large gas reservoirs at the
centers of galaxies, fuelling high star formation rates there.  As in
other barred galaxies, the Milky Way (MW)'s bar funnels gas inwards
\citep{binney+91, weiner_sellwood99, fux99}.  This gas gives rise to
structures such as the Central Molecular Zone (CMZ), spanning
$-1\degrees \la l \la 1.5\degrees$.  The CMZ contains $5-10 \times
10^7 \Msun$ of molecular gas \citep{bally+87, gusten89}, driving a
star formation rate of $\sim 0.14 \Msun $yr$^{-1}$
\citep{wardle_yusef-zadeh08}.
A molecular gas disk extends across $|l| < 6\degrees$ and $|b| <
1.6\degrees$ \citep{boyce_cohen94, dame_thaddeus94}.
\citet{liszt_burton80} and \citet{ferriere+07} interpreted the
observed molecular, atomic and ionized gas outside the CMZ to Galactic
longitude $|l| \sim 10 \degrees$ as a (tilted) disk with semi-major
axis of radius $\sim 1.4 \kpc$ with a hole at its center.  In external
galaxies, star formation in nuclear rings builds nuclear disks
\citep{kormendy_kennicutt04}.  In this Letter we demonstrate that the
high-\vgsr\ peaks in the line of sight velocity distributions (LOSVDs)
are consistent with the presence of a nuclear disk in the MW.

%
%

\section{Simulation}
\label{sec:simulation}

Here we use a high resolution simulation, with gas and star formation,
which develops a bar, driving gas to the center and forming a stellar
nuclear disk \citep{cole+14}, to derive the kinematic signatures of
such a disk.  We use these to guide our interpretation of the APOGEE
Data Release 12 \citep{alam+15} stellar velocity data for the inner
MW.  While the simulation was not designed to match the MW,
\citet{cole+14} showed that the nuclear disk that it forms is
qualitatively similar to those in external galaxies.

The simulation was evolved with the $N$-body$+$smoothed particle
hydrodynamics code {\sc gasoline} \citep{gasoline}.  The galaxy forms
out of gas cooling off a hot corona in pressure equilibrium within a
dark matter halo of virial mass $M_{200} = 9 \times 10^{11} \Msun$.
Both the dark matter halo and the initial gas corona are represented
by $5 \times 10^6$ particles.  As the gas cools and reaches high
density, star formation is triggered.  Star particles then provide
feedback via winds from massive stars, and types Ia and II supernovae
\citep{stinson+06}.  Gas particles all have initial mass of $2.7
\times 10^4 \Msun$ and star particles are spawned from gas with $35\%$
of this mass.  This high mass resolution allows us to use a high star
formation threshold of 100 $\mathrm{cm}^{-3}$ for the gas
\citep{governato+10}.  By the end of the simulation the galaxy has a
stellar mass of $6.5 \times 10^{10} \Msun$ in $\sim 1.1\times 10^7$
particles.  This large number of star particles provides a very fine
sampling of the mass distribution at the center of the model.  Further
details of the simulation are provided in \citet{cole+14}

The bar forms at around 4 Gyr.  After 6 Gyr a prominent nuclear disk
starts to form which, by 10 Gyr, has a semi-major axis of $1.5$ \kpc.
The nuclear disk is perpendicular to the bar and its stellar streaming
is perpendicular to the bar's.  At 10 Gyr the nuclear disk in the
simulation is quite massive and is thus unlikely to match any nuclear
disk in the MW.  Therefore here we consider the model at two earlier
times: at $\tm{1}=6$~Gyr, before the nuclear disk forms, and at
$\tm{2}=7.5$~Gyr when a strong nuclear disk is established.  Aside
from the nuclear disk becoming more massive and the bar growing
longer, the model at 10~Gyr is not qualitatively different from at
\tm{2}.

\subsection{Scaling to the MW and Viewing Perspective}

In order to compare to the MW, we rescale the model in both size and
velocity.  Size rescaling is accomplished by matching the size of the
bar to that of the MW.  Between \tm{1}\ and \tm{2}\ the average size
of the bar in the simulation, as measured from the radius at which the
phase of the $m=2$ Fourier moment deviates from a constant by more
than $10\degrees$ \citep{aguerri+03}, is 2.1~kpc.  Assuming that the
MW's bar has a semi-major axis of $3.5 \kpc$ \citep{gerhard02}, we
scale all coordinates by a factor of 1.67.  (Scaling to the more
up-to-date bar size of \citet{wegg+15}, $5 \kpc$, leads to a nuclear
disk which is much too large; because we seek a closer nuclear disk
size match, we scale to the older bar size, but this is not to imply
that the real MW bar semi-major axis is closer to $3.5\kpc$ than
$5\kpc$.)  The velocity scale factor is obtained by a least-squares
fit to the line of sight velocity dispersion of the model to
Abundances and Radial velocity Galactic Origins Survey
\citep{ness+13b} data for all stars within Galactocentric radius
$R_{\rm GC} < 3.5$ \kpc\ at $b$ = $5\degrees$, $7.5\degrees$ and
$10\degrees$ across $|l| < 15\degrees$.  We obtain a velocity scaling
factor of 0.48.  While these scalings lead to a model of roughly the
right size and rotational velocity we stress that the model still does
not match the MW and we only use it to qualitatively predict the
expected trends in the MW, not their magnitude or precise location.

We assume that the Sun is 8 \kpc\ from the Galactic Center, and place
the observer at $y=-8$ \kpc.  We orient the bar at $27\degrees$ to the
line of sight \citep{wegg+gerhard13}.  Since we compare our model with
APOGEE \citep{alam+15} data, which targets bright red giant stars, we
adopt a uniform selection function for star particles at $2 \kpc \leq
R_s \leq 10 \kpc$, where $R_s$ is the distance from the Sun
\citep{schultheis+14, hayden+15}.  Reducing the maximum $R_s$ to 8 kpc
does not significantly alter our conclusions.  We use an opening angle
of $0.5\degrees$ for each LOSVD, to match the size of the smallest
APOGEE bulge fields.  The (off-plane) line of sight with the least
particles contains over 2800 star particles while the best sampled
(mid-plane) field has over 57,000 star particles; thus the shapes of
the model LOSVDs are well determined.  The top row of Figure
\ref{fig:losvds} shows the model's surface density distribution.

\subsection{Line of Sight Velocity Distributions}

\begin{figure*}
\centerline{
\includegraphics[angle=-90.,width=0.5\hsize]{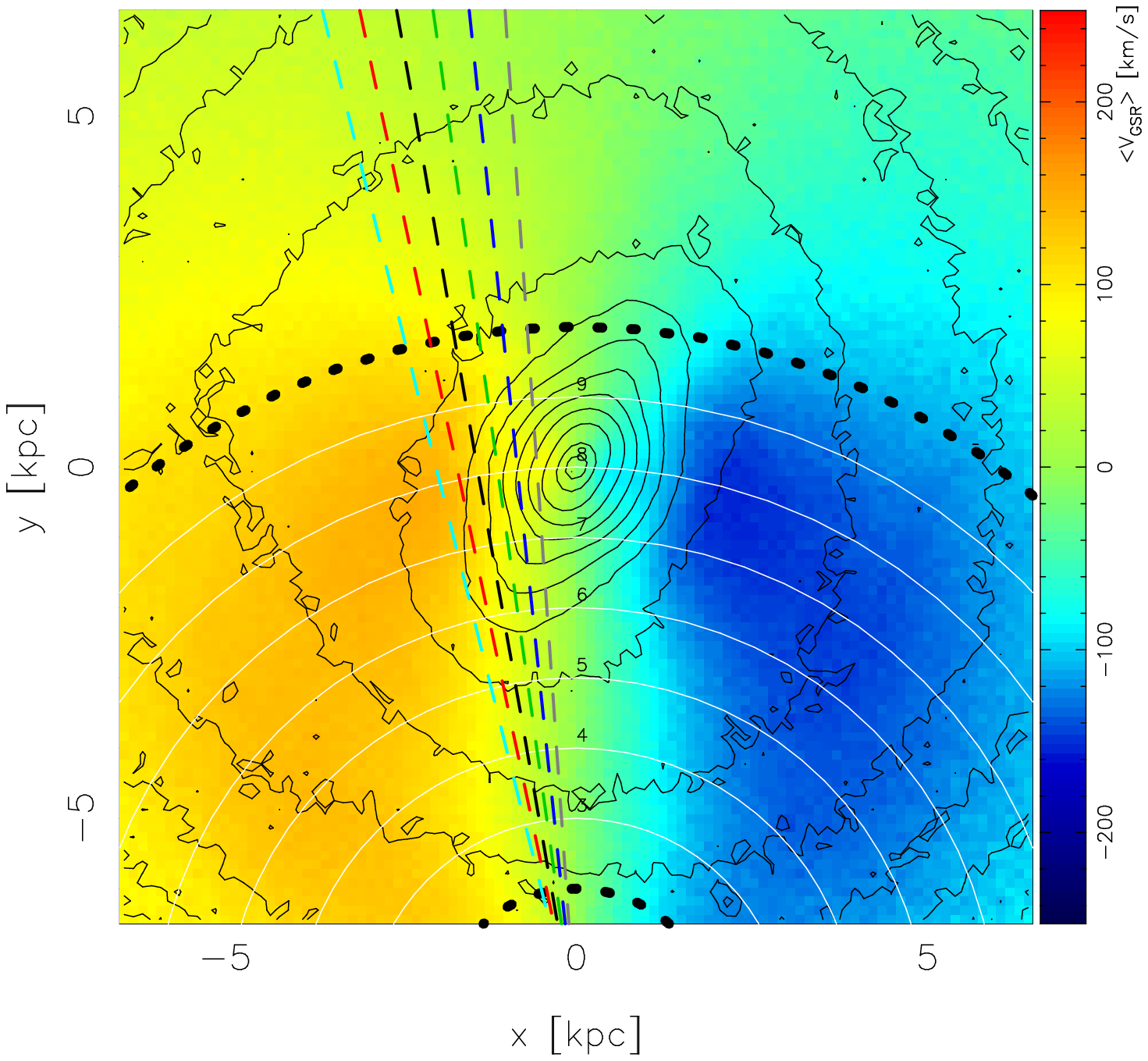} 
\includegraphics[angle=-90.,width=0.5\hsize]{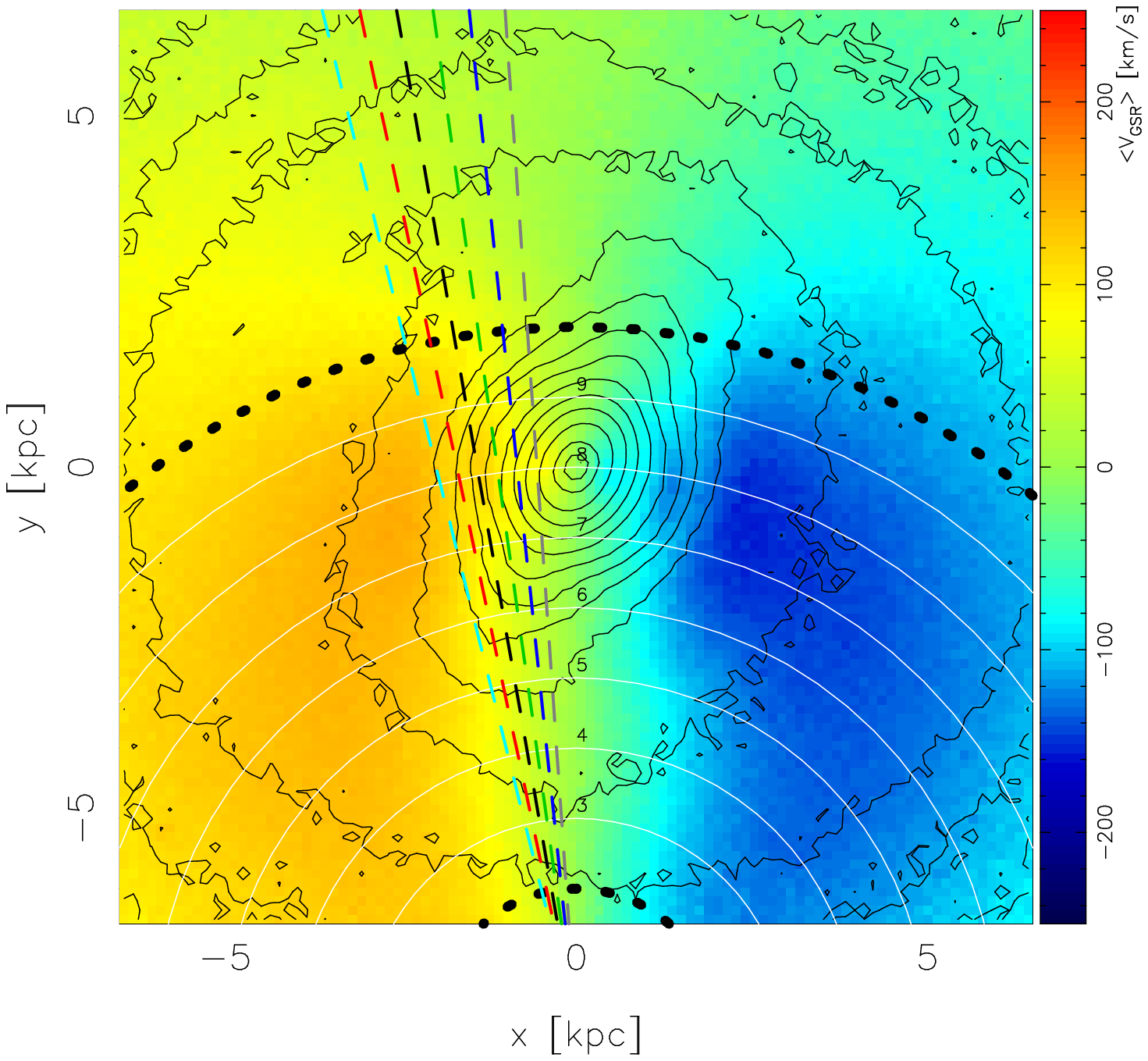} 
}
\centerline{
\includegraphics[angle=-90.,width=0.5\hsize]{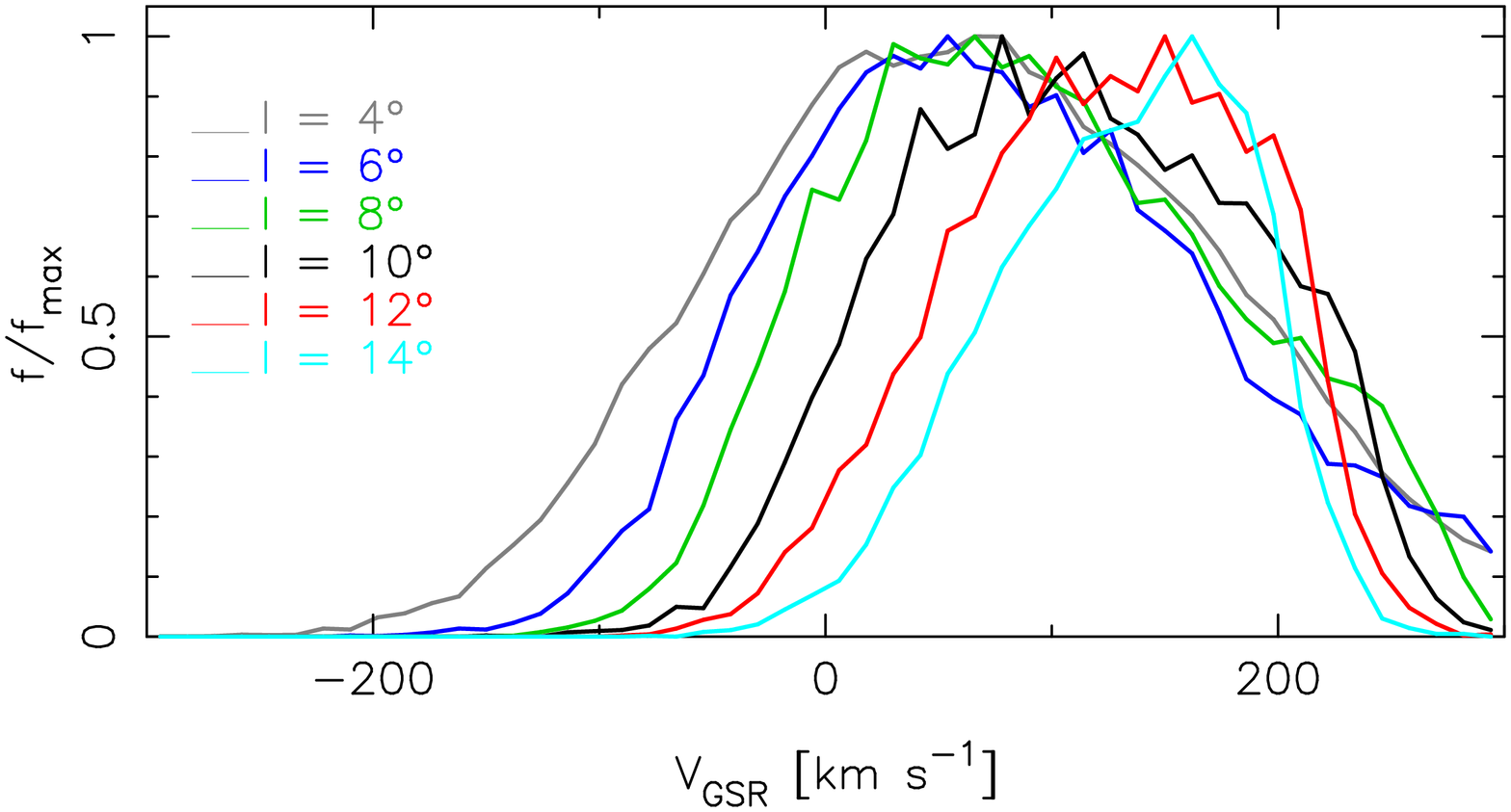}
\includegraphics[angle=-90.,width=0.5\hsize]{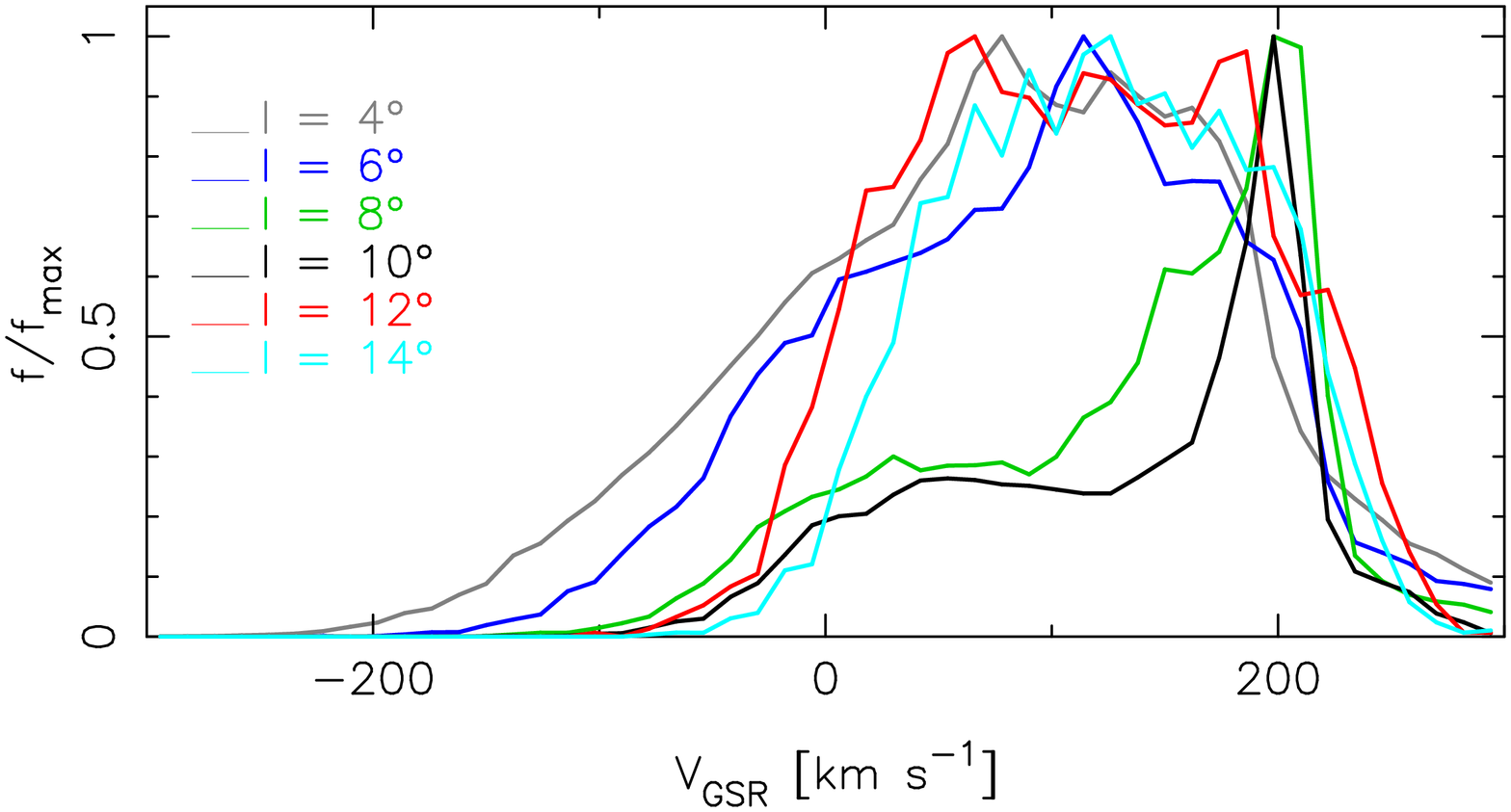}
}
\centerline{
\includegraphics[angle=-90.,width=0.5\hsize]{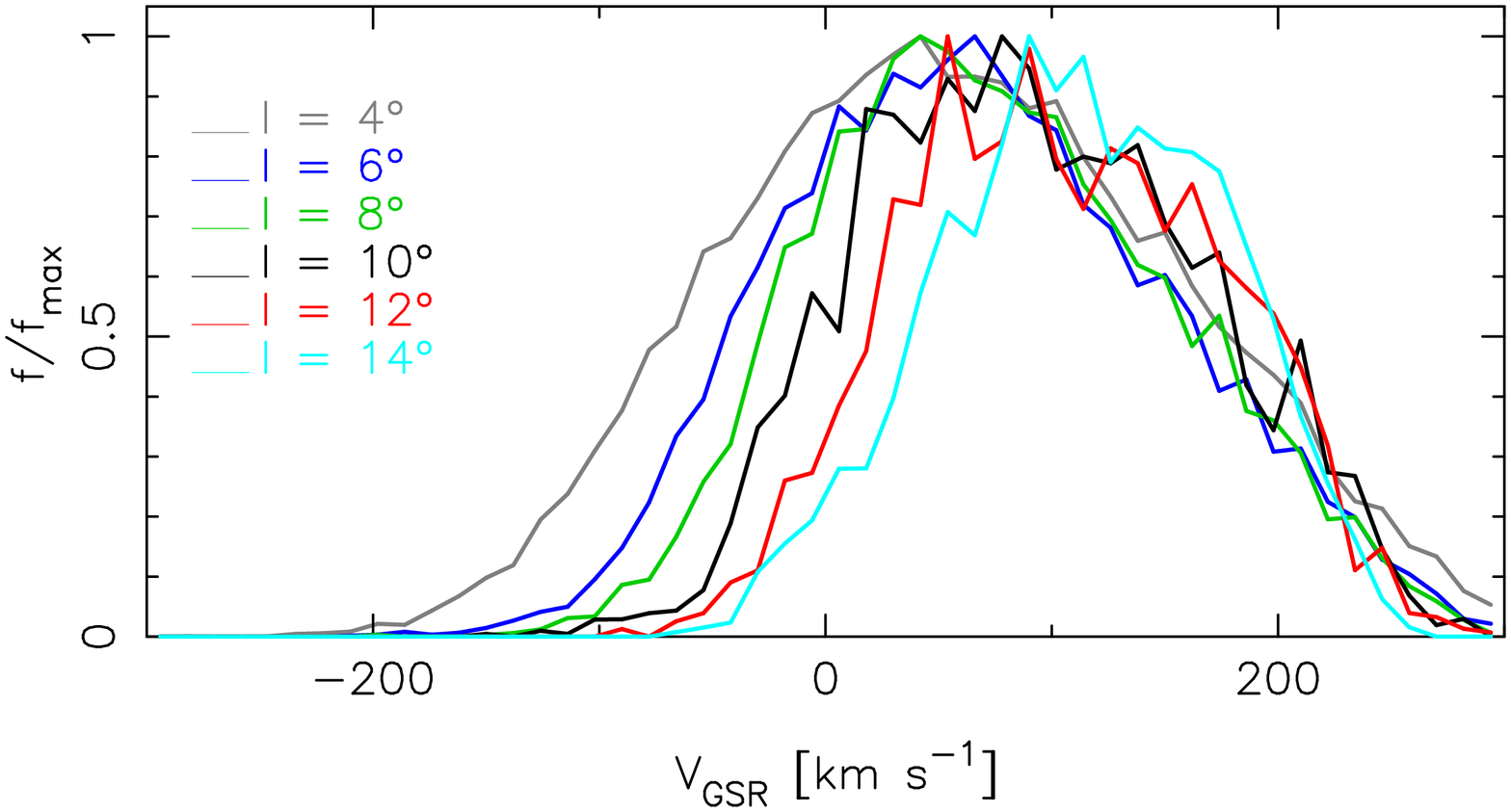}
\includegraphics[angle=-90.,width=0.5\hsize]{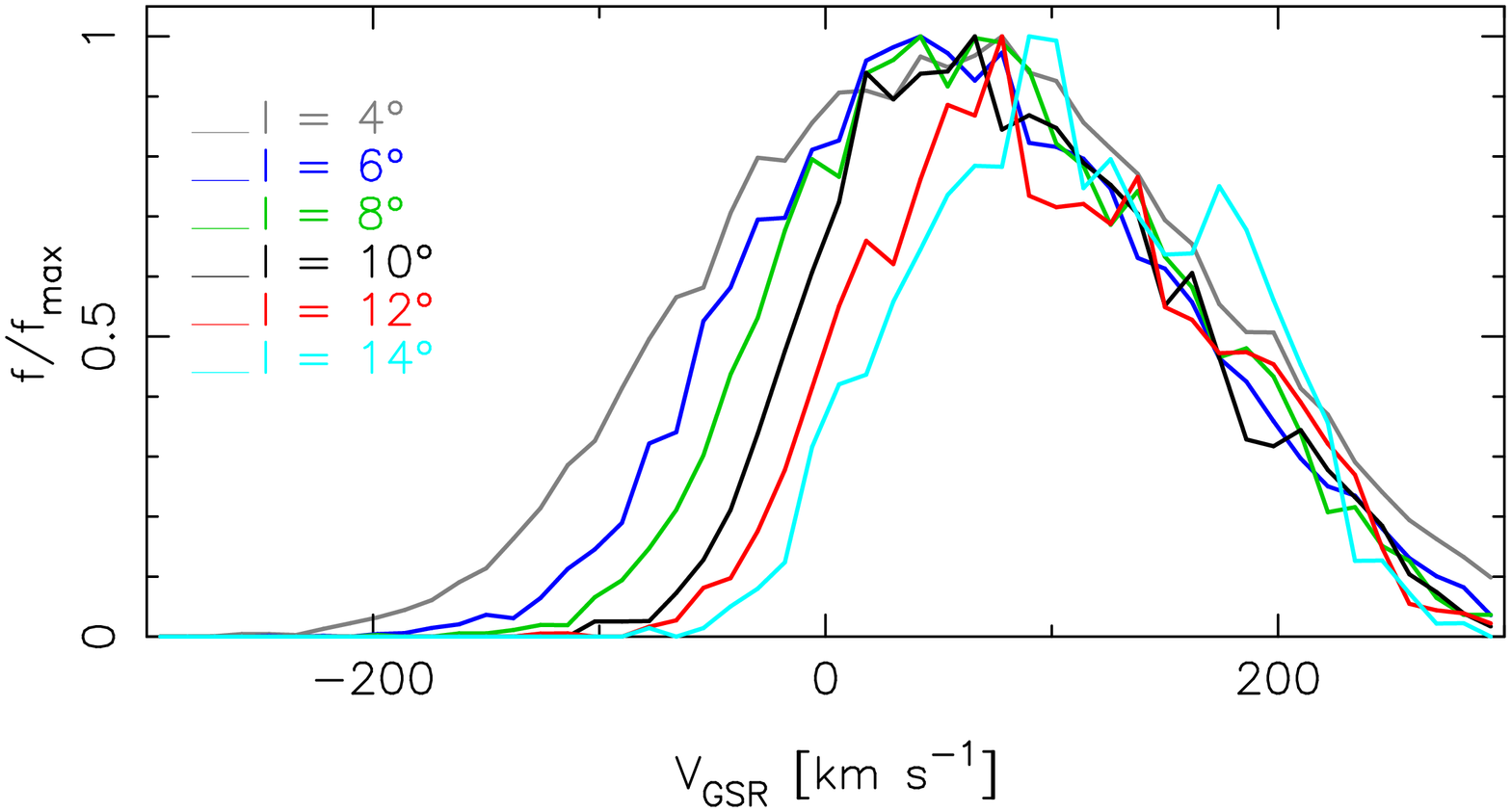}
}
\centerline{
\includegraphics[angle=-90.,width=0.5\hsize]{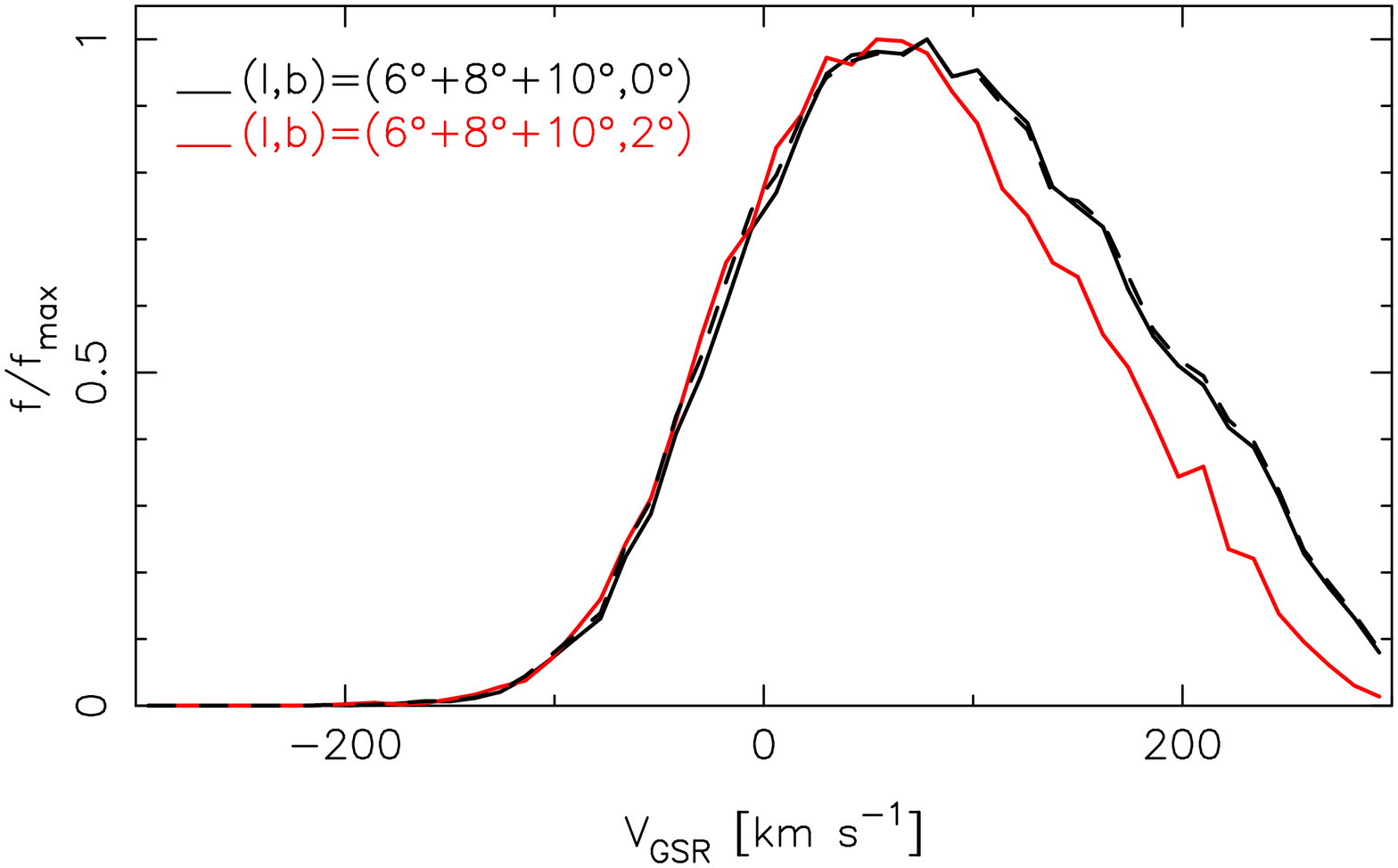}
\includegraphics[angle=-90.,width=0.5\hsize]{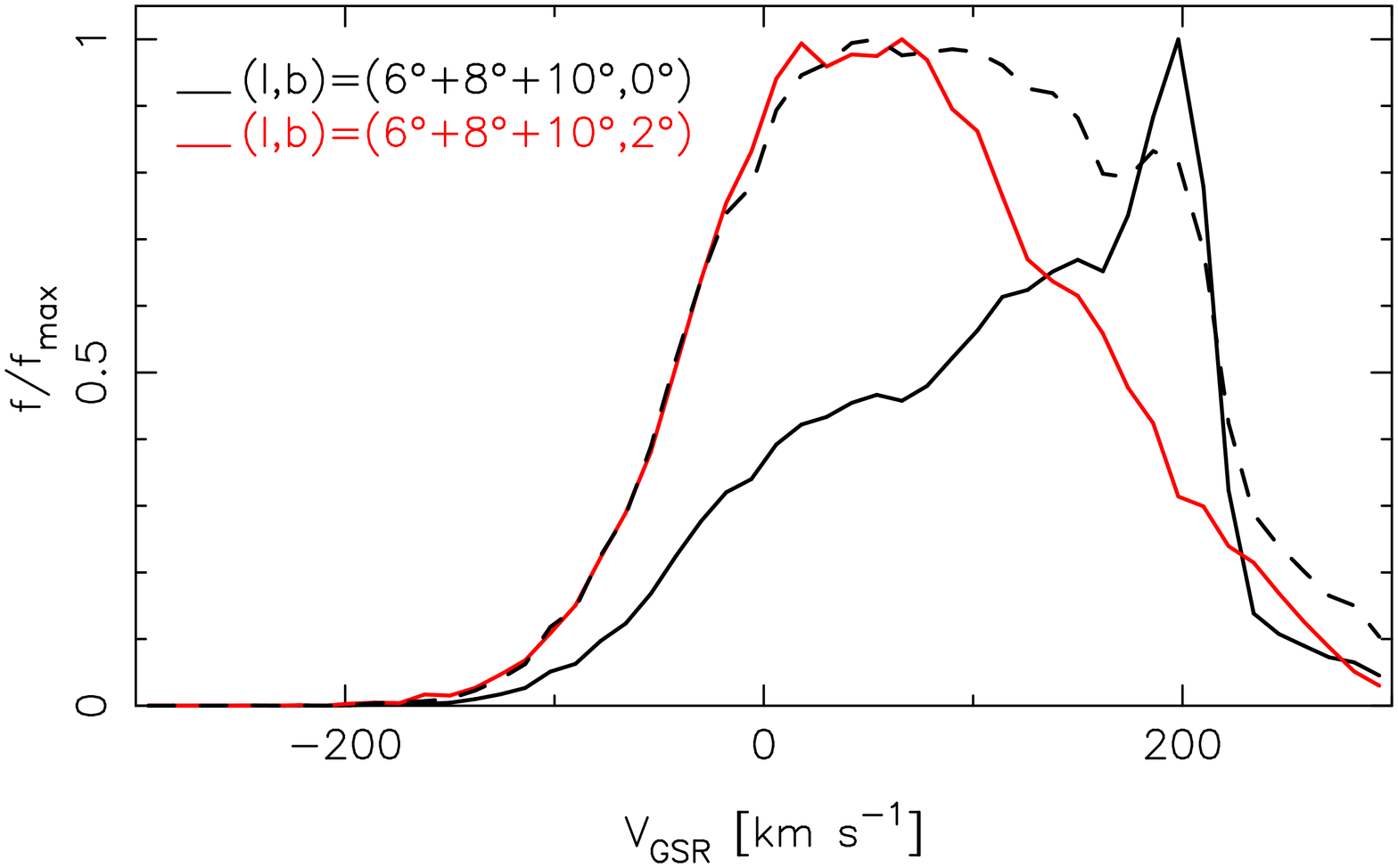}
}
\caption{Top row: Face-on views of the model: contours indicate the
  surface density while colors show \avg{\vgsr}.  The bold dotted
  circles indicate the radii between which star particles are chosen
  (the selection function).  The dashed lines show longitudes
  $4\degrees$-$14\degrees$ in $2\degrees$ steps, color-coded as in
  the next two rows.  Second row: Mid-plane LOSVDs for the different
  longitudes.  Third row: LOSVDs at $b=2\degrees$, colored as in the
  rows above.  Bottom row: Stacked model LOSVDs from $l=6\degrees$,
  $l=8\degrees$ and $l=10\degrees$ in the mid-plane (black) and at
  $b=2\degrees$ (red).  The dashed black lines show the effect of
  reducing the weights of star particles younger than 1 \Gyr\ by a
  factor of 5, to compensate for the very high star formation rate in
  the model.  All LOSVDs have been normalized to unit peak.  The left
  panels are at \tm{1}\ while the right ones are at \tm{2}.  }
\label{fig:losvds}
\end{figure*}

Viewing the model from the Solar perspective, we measure the
distribution of line of sight velocities in the Galactocentric
restframe, \vgsr.  Figure \ref{fig:losvds} shows the LOSVDs for
various lines-of-sight (indicated in the top row) in the mid-plane
($b=0\degrees$, second row) and off-plane ($b=2\degrees$, third row).
At \tm{1}\ each LOSVD at $l\leq 12\degrees$ has a single peak, both in
the mid-plane and off the plane.  The LOSVDs have a shoulder to high
\vgsr, which \citet{li+14} showed is produced by stars at large
distances seen close to tangentially.  The peak in \vgsr\ moves to
larger velocities with increasing $l$, but remains well below the
Galaxy's circular velocity.  By \tm{2}\ the LOSVDs at $l = 8\degrees$
and $l = 10\degrees$ have developed a second, high-\vgsr\ peak.  This
peak is more prominent than the low-\vgsr\ peak, due to the model's
very vigorous star formation in the nuclear disk, roughly ten times
higher than in the MW for the corresponding region.  This very high
star formation rate quickly leads to a relatively massive nuclear
disk; thus the relative amplitudes of the low- and high-\vgsr\ peaks
are {\it not} predictions of the model.  Indeed if we reduce the
weight of star particles younger than 1\Gyr\ by a factor of 5, to
compensate for the high star formation rate of the model, then the
high-\vgsr\ peaks become smaller than the main peaks, as seen in
Figure \ref{fig:losvds}.  The distribution around the high-\vgsr\ peak
is narrower (\ie\ cooler) than that around the main peak and is skewed
toward low \vgsr.  Interior to $l = 8\degrees$, the LOSVDs are
broadened relative to those at \tm{1}, but no high-\vgsr\ peak is
evident.  At $l \geq 14\degrees$ no high-\vgsr\ peak is present in the
mid-plane, indicating that the structure responsible for the feature
does not extend this far.  The off-plane and mid-plane LOSVDs are not
substantially different at \tm{1}, aside from the mid-plane hosting
more stars at $\vgsr \geq 100~\kms$.  At \tm{2} the high-\vgsr\ peaks,
which dominate the mid-plane, are entirely absent in the $b =
2\degrees$ LOSVDs.  Therefore the presence of a nuclear disk is only
evident in the mid-plane.  As in the MW, outside the nuclear disk, the
off-plane LOSVDs at $(l,b) = (14\degrees,2\degrees)$ also contain a
statistically significant high-\vgsr\ peak/shoulder, but this is also
present at \tm{1}, and is not related to the nuclear disk.  Thus the
kinematic signatures of a nuclear disk are (1) a second,
high-\vgsr\ peak at roughly the circular velocity, (2) which is absent
a few degrees off the mid-plane, (3) is kinematically cooler than the
low-\vgsr\ peak, and (4) is skewed toward low \vgsr.

\subsection{LOSVD Stacking}

The top row of Figure \ref{fig:losvds} shows color-coded maps of the
average \vgsr, \avg{\vgsr}; the peak velocities at orbit tangent
points manifest as the characteristic ``winged'' pattern of the
\avg{\vgsr}\ fields.  Although the two \avg{\vgsr}\ maps show the
model before and after the nuclear disk forms, they are not very
different, indicating that the formation of the nuclear disk does not
lead to a wholesale change of the galaxy as much as populating new
parts of its phase space.  At the low longitudes of the nuclear disk,
large \avg{\vgsr} occurs only close to the galactic center while at
other radii \avg{\vgsr} is smaller.

Even with a survey the size of APOGEE, the number of stars in
individual fields is still relatively small, giving a low
signal-to-noise ratio for any second peak in any one field
\citep{li+14}.  In order to overcome this difficulty, we note that the
\vgsr\ of the second peak does not change significantly with longitude
at $6\degrees \leq l \leq 10\degrees$.  Therefore by stacking the
LOSVDs we can enhance the signal-to-noise ratio of the
high-\vgsr\ peak.  Because the main peak is dominated by stars
streaming along the bar, and \avg{\vgsr}\ of these changes with $l$,
the main peak in a stacked LOSVD will be quite broad.  If we include
$l < 4\degrees$, then the exponentially higher density of disk and bar
stars near the center masks out any features at high \vgsr.  In the
bottom panels of Figure \ref{fig:losvds} we present a stack of the
model's LOSVDs at $l=6\degrees$, $8\degrees$ and $10\degrees$.  As
with the individual LOSVDs, a peak at high \vgsr\ is evident at
\tm{2}\ in the mid-plane but is absent at $b=2\degrees$.  Moreover
this second peak is still cooler than the low-\vgsr\ peak, and remains
skewed toward it.  Thus stacking LOSVDs preserves the kinematic
signatures of a nuclear disk, and provides a reliable method for
searching for a nuclear disk in the APOGEE data.

%
%

\section{APOGEE data}
\label{sec:apogee}

\subsection{Data Selection}

We select APOGEE survey stars in the fields of interest, excluding
stars with the STAR BAD flag (corresponding to poor stellar parameter
fits) and those flagged as flux and telluric standards.  Stars with a
velocity scatter between different visits of more than $5 \kms$ are
also removed. (The same analysis including also stars flagged as STAR
BAD, which leads to 763 in the plane and 1401 out of the plane, gives
results in agreement with the more conservative cut.)

The small numbers of stars in the APOGEE commissioning data resulted
in peaks with low signal-to-noise ratio.  We increase the statistical
significance of a high-\vgsr\ peak by stacking the APOGEE DR12 data in
the longitude range $6\degrees \leq l \leq 8\degrees$ for fields in
the mid-plane and off-plane at $|b| = 2\degrees$ (totalling 617 and
1114 stars, respectively).  Table \ref{tab:fields} lists the fields
stacked together and the number of stars used from each field.

\begin{centering}
\begin{table}
\vbox{\hfil
\begin{tabular}{ccccc}\hline
\multicolumn{1}{c}{Field} &
\multicolumn{1}{c}{$l$ [$\degrees$]} &
\multicolumn{1}{c}{$b$ [$\degrees$]} &
\multicolumn{1}{c}{$N_*$} &
\multicolumn{1}{c}{Stack} \\

4336 & 6.0 & 0.0 & 471 & mid-plane \\
4355 & 8.0 & 0.0 & 146 & mid-plane \\
4365 & 5.7 & 2.0 & 387 & off-plane \\
4366 & 5.7 & -2.0 & 424 & off-plane \\
4373 & 7.8 & -2.0 & 154 & off-plane \\
4377 & 7.7 & 2.0 & 149 & off-plane \\ \hline

\end{tabular}
\hfil}
\caption{APOGEE fields used to construct the mid-plane and off-plane
  stacks.  $l$ and $b$ are the Galactic longitude and latitude,
  respectively, of the field centers.  $N_*$ is the number of stars
  selected in each field.  }
\label{tab:fields}
\end{table}
\end{centering}

\begin{figure}
\includegraphics[angle=0.,width=\hsize]{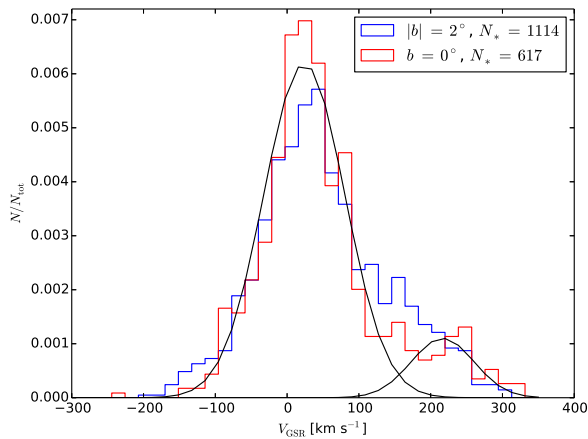}
\caption{The stacked APOGEE LOSVDs for the mid-plane (red histogram)
  and at $|b|=2\degrees$ (blue histogram).  The fields used are listed
  in Table 1.  The black lines show the two Gaussians fitted to the
  mid-plane LOSVD.}
\label{fig:apogeevgsrstack}
\end{figure}

Figure \ref{fig:apogeevgsrstack} plots these two stacked APOGEE
LOSVDs.  The mid-plane stack has a clear second peak at $\vgsr \sim
220 \kms$, corresponding to roughly the circular velocity of the MW in
the bulge region \citep{sofue+09}.  No comparable second peak is
visible in the off-plane stacked LOSVD, which is non-Gaussian and
skewed toward high \vgsr, \ie\ it has a shoulder to high
\vgsr\ \citep{li+14}.  A Kolmogorov-Smirnov test shows that the null
hypothesis that the mid-plane and off-plane LOSVDs are drawn from the
same distribution has a relatively low $p$-value of $0.04$.

We fit two Gaussians to the mid-plane stacked LOSVD in the range $-300
\kms \leq \vgsr\ \leq 300 \kms$, constrained such that the smaller
Gaussian contains less than $25\%$ of the stars (to avoid fitting just
the skewed low-\vgsr\ distribution with two Gaussians).  We obtain a
low-\vgsr\ component having mean velocity $\avg{\vgsr} = 24 \kms$ and
standard deviation $\sgsr = 57 \kms$, while the high-\vgsr\ component
has $\avg{\vgsr} = 217 \kms$ and $\sgsr = 44 \kms$, making it cooler
than the low-\vgsr\ component.  These two Gaussians are also shown in
Figure \ref{fig:apogeevgsrstack}.  The velocity distribution at $\vgsr
\geq 200 \kms$ hints at a skewness opposite to that of the main
distribution, but the signal-to-noise ratio is still too low for a
robust measurement.

\begin{figure}
\includegraphics[angle=0.,width=\hsize]{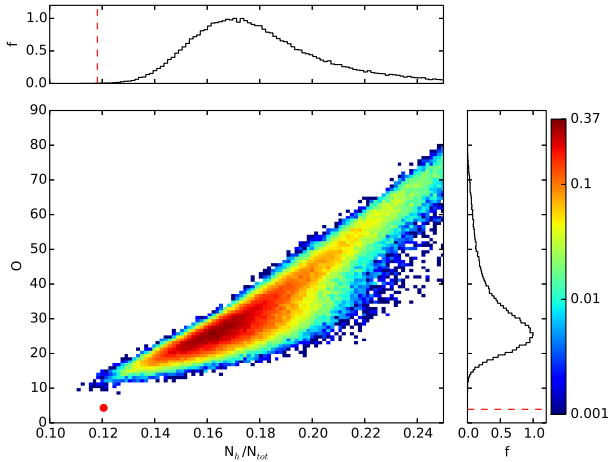}
\caption{Frequency distribution of properties of double-Gaussian fits
  for the off-plane APOGEE stacked LOSVD sub-sampled $N_s = 10^5$
  times to 628 stars.  The side-panels indicate the distributions over
  the individual variables, normalized to unit peak.  The parameters
  for the fit to the mid-plane APOGEE stack are indicated by the
  filled red circle.  In the side panels, the vertical dashed red
  lines indicate the values of $N_h/N_{\rm tot}$ and $O$ for the
  mid-plane stack.  }
\label{fig:statistics}
\end{figure}

The high-\vgsr\ Gaussian has a significant number of stars associated
with it, and is significantly separated from the low-\vgsr\ Gaussian.
In order to test the likelihood of such a second peak arising purely
from Poisson noise, we perform Monte-Carlo tests drawing 617 stars
from the off-plane stacked LOSVD.  Fitting two Gaussians as before to
the resulting LOSVD, we label as $G_l$ and $G_h$ the low- and
high-\vgsr\ components, respectively.  We repeat this procedure
100,000 times, and for each we compute $N_h/N_{\rm tot}$, the ratio of
stars in the high-\vgsr\ component to the total number of stars, and
the overlap of the two components, defined as
\begin{equation}
O = \int G_l G_h d\vgsr.  
\end{equation}
The results are presented in Figure \ref{fig:statistics}; the observed
mid-plane stacked LOSVD has $N_h/N_{\rm tot} = 0.12$ and $O = 4.3$.
Only $0.025\%$ of the Monte-Carlo samples have $N_h/N_{\rm tot} \leq
0.12$, while none of them have overlap $O \leq 8$, showing that the
observed double-peaked mid-plane stacked LOSVD is highly unlikely to
result from Poisson noise.
The APOGEE data therefore show a statistically significant
double-Gaussian LOSVD in the mid-plane, the properties of which agree
with 3 of the 4 kinematic signatures of a nuclear disk from the
simulation.  While the signal-to-noise is too low to be sure if the
high-\vgsr\ peak is skewed to low \vgsr, the data are suggestive that
it is.  Therefore a kiloparsec-scale nuclear disk can explain the
high-\vgsr\ peaks in the APOGEE data.

A simple estimate for the nuclear disk mass can be obtained from the
fraction of stars in the high-\vgsr\ component of the double-Gaussian
fit to the mid-plane LOSVD.  If we conservatively assume that the
nuclear disk mass contained within $|z| \leq 150\pc$ and $4\degrees
\leq |l| \leq 8\degrees$ is $12\%$ of the total mass of the
Besan\c{c}on Galaxy model \citep{robin+12} within this volume we
obtain a lower limit to the mass of the nuclear disk $\sim 5.8 \times
10^7\Msun$.

%
%

\section{Discussion}

Attempts to explain the high-\vgsr\ peak directly via collisionless
bar simulations fails \citep{nidever+12, li+14}.  However
\citet{molloy+15} demonstrated that resonant, bar-supporting 2:1 x1
(with some mixture of higher order resonance) orbits by themselves can
produce second peaks. Subsequently \citet{aumerschoenrich15} argued
that the selection function of APOGEE favors young stars recently
trapped into resonant orbits.  Their interpretation requires that the
stars in the high-\vgsr\ peaks are younger.
The other main 2:1 resonant orbit family of bars, the x2 family, is
orientated perpendicular to the bar.  This family is generally very
poorly populated in the absence of gas \citep{sparke_sellwood87,
  pfenniger_friedli91}, but when gas is present it is driven inwards
by the bar and settles into x2 orbits \citep{binney+91}.  The gas can
then form stars and produce nuclear rings and disks.  We propose that
the high-\vgsr\ peak corresponds to a kiloparsec-scale disk composed
of stars on x2 orbits.  These orbits are stable and therefore our
model does not require that the stars in the high-\vgsr\ peak are
young.

Nuclear disks are known in many external galaxies
\citep{scorza+vdbosch98, zasov+moiseev99, pizzella+02, emsellem+04,
  krajnovic+08, ledo+10}; the presence of one in the MW is
therefore not unusual.
Nor is the kiloparsec scale unusual as a fraction of the bar size.
For instance in NGC~3945 the ratio of semi-major axes of the nuclear
disk to bar is $\sim 0.15-0.18$ \citep{erwin_sparke99, cole+14},
whereas for the MW this ratio is $\sim 0.2$, if we adopt
\citet{wegg+15}'s $5 \kpc$ bar.  The gas ring in the simulation is
$\sim 5\times$ larger than the MW's CMZ, which is coincident with a
stellar disk \citep{launhardt+02, schoenrich+15}.  The large size of
the gas ring in the model is a consequence of the still low resolution
($50 \pc$) of our simulation \citep{lizhi+15, sormani+15}.  This
difference implies that the nuclear disk in the MW is not currently
forming stars across its full extent.

We anticipate that this proposal will inspire further detailed mapping
of the central mid-plane of the MW.  We will provide predictions from
our model of a kiloparsec-scale nuclear disk elsewhere.

\bibliographystyle{aj.bst}

\acknowledgements V. P. D. is supported by STFC Consolidated grants
\#~ST/J001341/1 and \#~ST/M000877/1, while D. R. C. is supported by STFC
Consolidated grant \#~ST/J001341/1.  M. N. is funded by the European
Research Council under the European Union's Seventh Framework
Programme (FP 7) ERC Grant Agreement \#~321035.  The authors thank the
ESF GREAT programme for funding which has supported this research.
The simulation used in this paper was run at the High Performance
Computing Facility of the University of Central Lancashire.  We thank
the anonymous referee for a very thoughtful report that helped
substantially improve this paper. Funding for SDSS-III has been
provided by the Alfred P. Sloan Foundation, the Participating
Institutions, the National Science Foundation, and the U.S. Department
of Energy Office of Science. The SDSS-III web site is
http://www.sdss3.org/. SDSS-III is managed by the Astrophysical
Research Consortium for the Participating Institutions of the SDSS-III
Collaboration including the University of Arizona, the Brazilian
Participation Group, Brookhaven National Laboratory, Carnegie Mellon
University, University of Florida, the French Participation Group, the
German Participation Group, Harvard University, the Instituto de
Astrofisica de Canarias, the Michigan State/Notre Dame/JINA
Participation Group, Johns Hopkins University, Lawrence Berkeley
National Laboratory, Max Planck Institute for Astrophysics, Max Planck
Institute for Extraterrestrial Physics, New Mexico State University,
New York University, Ohio State University, Pennsylvania State
University, University of Portsmouth, Princeton University, the
Spanish Participation Group, University of Tokyo, University of Utah,
Vanderbilt University, University of Virginia, University of
Washington, and Yale University.

\end{document}